\documentclass[11pt]{article} % use larger type; default would be 10pt

\usepackage[utf8]{inputenc} % set input encoding (not needed with XeLaTeX)

\usepackage{geometry} % to change the page dimensions
\usepackage{booktabs} % for much better looking tables
\usepackage{indentfirst}
\usepackage{array} % for better arrays (eg matrices) in maths
\usepackage{paralist} % very flexible & customisable lists (eg. enumerate/itemize, etc.)
\usepackage{verbatim} % adds environment for commenting out blocks of text & for better verbatim
\usepackage{fancyhdr} % This should be set AFTER setting up the page geometry
\usepackage{amssymb}
\usepackage{multicol}
\usepackage{graphicx}
\usepackage{balance}
\usepackage{color}
\usepackage{xcolor}
\usepackage{multirow}
\usepackage{caption}
\usepackage{float}
\usepackage{subfigure}
\restylefloat{table}
\newtheorem{problem}{Problem}
\usepackage{epsfig,amsfonts,amsmath,cases, latexsym, times, algorithmic, algorithm, bbm, soul,wrapfig,comment}
\providecommand{\keywords}[1]{\textbf{\textit{Index terms---}} #1}

\begin{document} 
%\begin{frontmatter}
\title{\line(1,0){250} \line(1,0){250}\\ Semi-Blind and \emph{$l_{1}$} Robust System Identification for Anemia Management}
\author{Affan Affan\thanks{Affan Affan is Ph.D student at the Department of Electrical and Computer Engineering, University of Louisville,
Louisville, KY, 40292 USA.}, Tamer Inanc\thanks{Tamer Inanc is the Associate Professor at the Department of Electrical and Computer Engineering, University of Louisville, Louisville, KY, 40292 USA.}\\
Email: affan.affan@louisville.edu, t.inanc@louisville.edu\\Electrical and Computer Engineering Department, J.B Speed School of Engineering, \\University of Louisville, United States\\ \line(1,0){250}\line(1,0){250} }
\pagenumbering{gobble}
\maketitle
\begin{abstract}
Chronic diseases such as cancer, diabetes, heart diseases, chronic kidney disease (CKD) require a drug management system that ensures a stable and robust output of the patient's condition in response to drug dosage. In the case of CKD, the patients suffer from the deficiency of red blood cell count and external human recombinant erythropoietin (EPO) is required to maintain healthy levels of hemoglobin (Hb). Anemia is a common comorbidity in patients with CKD. For an efficient and robust anemia management system for CKD patients instead of traditional population-based approaches, individualized patient-specific approaches are needed. Hence, individualized system (patient) models for patient-specific drug-dose responses are required. In this research, system identification for CKD is performed for individual patients. For control-oriented system identification, two robust identification techniques are applied: (1) \emph{$l_{1}$} robust identification considering zero initial conditions and (2) semi-blind robust system identification considering non-zero initial conditions. The EPO data of patients are used as the input and Hb data is used as the output of the system. For this study, individualized patient models are developed by using patient-specific data. The ARX one-step-ahead prediction technique is used for model validation at real patient data. The performance of these two techniques is compared by calculating minimum means square error (MMSE). By comparison, we show that the semi-blind robust identification technique gives better results as compared to \emph{$l_{1}$} robust identification.

\end{abstract}

\keywords{\emph{$l_{1}$} robust system identification, Semi-blind robust system identification, Robust system identification, Anemia management, Chronic kidney disease.}

%\end{frontmatter}

\section{Introduction}
  \label{sec:intro}
In recent years, the need for drug management in bio-engineering is becoming one of the prominent applications for control systems researchers. For feedback control systems, the mathematical model of the plant is important to apply control and prediction algorithms. As the system model for the individual patient is different and complex to write its mathematical form. In chronic kidney disease (CKD), the human body is not able to produce enough erythropoietin, a glycoprotein that produces red blood cells~\cite{adamson1968erythropoietin}. The patients of this disease are treated with the recombinant human erythropoietin (EPO). The dosage of EPO to the human body increases the hemoglobin (Hb) levels~\cite{eschbach1985anemia}. According to the National Kidney Foundation’s Dialysis, the Hb level should be in the range of 11 and 12 $g/dL$ in the response of EPO dosage~\cite{valderrabano1996erythropoietin}. 

Presently, the medical centers have developed their own EPO management systems based on the average (population-based) response medication. While population models may be useful in the analysis of drug properties at large, they are not well suited to guide treatment of individual patients. In order to effectively address the impact of inter- and intra-individual variability in dose-response characteristics, personalized, patient-specific models are needed.
Therefore, medical institutes are in need of a stable and robust system for individualized anemia management. The key point in designing an efficient individualized anemia management system is to find accurate patient-specific dose-response models. 

Many researchers have attempted to develop patient models for anemia management. Bayesian-based drug delivery using population patient data is discussed in~\cite{bellazzi1993drug}. Artificial Intelligence-based neural network models are discussed in~\cite{gaweda2003pharmacodynamic}~\cite{guerrero2003use}~\cite{gaweda2009improving}. Some researchers attempted to identify individual patient models in~\cite{muezzinoglu2006approximate}. In~\cite{martin2003dosage} the main focus was to predict the value for EPO instead of Hb, which is not desired as predicting Hb level gives values of EPO but not true for vise versa.  However, 

most of these models are based on predetermined model structure and noise distribution, which is not suitable for anemia management as each patient poses different model characteristics. The models obtained by the classical system identification techniques do not yield good results as it assumes that predefined model structure and model order are close to the actual system and one mathematical model works for all dynamics regardless of complexity and disturbances in the system. The modeling stage should include the effect of all disturbances and uncertainties which are being introduced during operations. In contrast to classical identification techniques, robust system identification takes into account system uncertainties, unmodeled dynamics and model complexity, i.e., there is no assumption on the model order, uncertainties and noise affecting data.

For robust control, the design requires the nominal model and uncertainty bound on this model~\cite{ljung1999system}. Therefore, robust system identification techniques should provide a nominal model and uncertainty bound on it~\cite{mazzaro2001robust}. The robust system identification procedure does not require large measurement data set or information of measurement noise nor the information on the structure of the system to be identified is required. In this technique, the information on the maximum gain of the system $K$, the stability margin of the system response $\rho$, and a bound on the noise is required. Depending on the nature of the a posteriori information, the robust identification techniques may lead to different identification methods. The selection of robust system identification method depends on the data type as time-domain data and frequency domain data pose different characteristics and different techniques are applied to obtain a model for the data type~\cite{schoukens2010use}. For frequency domain data, the \emph{$H_{\infty}$}-identification technique is implemented, which calculates the uncertainty bounds in terms of the \emph{$H_{\infty}$}-norm. For time-domain data, the \emph{$l_{1}$}-identification technique is implemented, which calculates an \emph{$l_{1}$}-error bound. However, sometimes both time and frequency domain data may be available for a single plant. In this case, the mixed \emph{$H_{\infty}$}/\emph{$l_{1}$} robust identification procedure is used~\cite{inanc1996mixed}.

In this research work, the individual patient model for anemia management is obtained using robust system identification techniques and patient-specific data. For system identification, the \emph{$l_{1}$} robust system identification and Semi-Blind system identification techniques~\cite{ma2006semi} are used. The system model is a combination of a nominal model and uncertainty bound. Both of these identification techniques, develop a nominal model and uncertainty bound using patient-specific dose-response (EPO-Hb) data.  

We summarize our contributions as follows:

\begin{enumerate}
\item We present an individualized patient model for anemia patients using \emph{$l_{1}$} robust system identification assuming zero initial conditions. The system model is reduced to 4th order which is suitable for control synthesis techniques. 

\item We present an individualized patient model for anemia patient using semi-blind robust system identification by incorporating the effects of initial conditions instead of assuming it zero. 

\item Validation of models obtained by \emph{$l_{1}$} robust system identification and semi-blind system identification using ARX one-step-ahead prediction~\cite{soderstrom1997least}. 

\item Comparison and error analysis of the obtained models for prediction are also provided using minimum mean squared error (MMSE) analysis.
\end{enumerate}

The remainder of this paper is organized as follows. First, we discuss the \emph{$l_{1}$} robust system identification in Section~\ref{sec:l1}. Then, we introduce the semi-blind robust system identification in Section~\ref{sec:semi}. Finally, we present the system models results obtained by the one-step ahead prediction along with the error analysis for both techniques in Section~\ref{sec:ARX} followed by the conclusion.   

\section{System Identification}
\begin{figure}[H] \centering
    \centering
    \includegraphics[height=5cm]{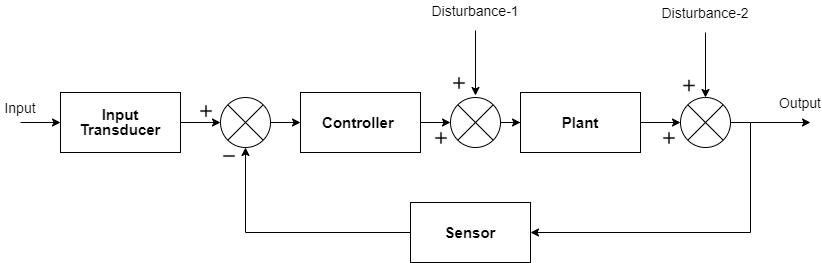}
    \caption{Feedback control system of a plant. }
    \label{fig:feedback}
\end{figure}

The plant model is the set of mathematical equations that describes the system behavior in response to the system input. The typical feedback control system is shown in Fig.~\ref{fig:feedback}.
The purpose of system identification is to develop a low order model of the plant by using finite, noisy sensor data, which can be controllable and observable by implementing suitable controller and observer respectively. In this work, the system model is the combination of a parametric and non-parametric portion given  as:
\begin{equation}
		G=G_{p}+G_{np}
\end{equation}
The non-parametric portion $G_{np}$ describes the internal behavior of the plant. This portion requires less prior knowledge of the system such as a model for internal behavior of physiological systems. On the other hand, the parametric portion $G_p$ describes the input to output relation. For the parametric portion, the prior knowledge is important, which urges the need for initial condition effect. The robust system identification techniques are good for non-parametric portions such as \emph{$l_{1}$} robust identification. To incorporate the effect of initial conditions, the semi-blind identification technique is introduced, which uses the same framework as \emph{$l_{1}$} robust identification but also computes the parametric portion. \\
Next, we formally define the problem. 
\begin{problem}
Considering equation~\ref{eq:pro}, obtain the plant model $G$ which gives output $y_k$ as Hb level in response to input $u_k$ of EPO dosage to patients of CKD.
\end{problem}
  \begin{align}
    y_i=Gu_i+\eta_i,\label{eq:pro}\\
    \left | \eta_{i} \right |\leq \epsilon_{i},   
   \end{align}
where $\epsilon_{i}$ is the maximum bound on noise, which is unknown but bounded.

\section{\emph{$l_{1}$} Robust Identification}
\label{sec:l1}
The \emph{$l_{1}$} robust identification, is the method to obtain a system model with a minimum prior knowledge about the system~\cite{dahleh1993sample}.  The classical identification techniques assume that the model has a fixed order and structure and probabilistic noise distribution. However, \emph{$l_{1}$} robust identification does not assume a fixed order model or noise distribution~\cite{jacobson1991worst}. It requires minimal prior information, such as the maximum gain on the system $K > 0$, the lower bound on the relative stability margin of the system dynamics $\rho > 1$, and an upper bound on the noise in measurements $\epsilon > 0$. The model obtained by this technique provides the bound on modeling error which shows the robustness of this method as shown in the following equation.
  \begin{align}
    G(z)=G_o(z) + \zeta,\label{eq:nominal}\\
    \left \| G(z)-G_o(z)  \right \|_{\infty} \leqslant \zeta \label{eq:bound},   
   \end{align}
where $G_o(z)$ is the nominal model and $\zeta$ is the mismatch between the nominal model and plant model which is the effect of model error and measurement error. The equation~\ref{eq:bound} provides the maximum bound on the modeling error. This bound is hardbound whereas the classical methods provide soft bounds based on probabilistic assumptions.    
By this method, consider a linear time-invariant (LTI) operator $g$ of system class \emph{$S$}, which maps $u$ from input space to $y$ in output space. The impulse response of $g$ is given as follows: 

\begin{equation}
    G(z)=\sum_{i=0}^{\infty} g_{i}z^{-i},
\end{equation}
where 
\begin{equation}
    g_{i}=K \rho_{i}, 
\end{equation}
where $K$ is the bound on impulse response and $\rho$ is the decay rate. By the discussion, for $N$ samples of data, the robust identification problem can be defined as follows:
\begin{problem}
Given the a priori and the a posteriori information, determine:

\begin{itemize}
    \item whether the priori and posteriori information are consistent, i.e decide whether the models in G$_{\infty , \rho}$ interpolates the given measurement points with error bounded by the priori error information. The priori and posteriori information is consistent if and only if, $\Gamma$ is a non-empty set.  
    \begin{equation}
        \Gamma \doteq g \in S |_{y_i}= (g*u_i)+ \eta_i
    \end{equation}
    \item If the two sources of information are consistent, then obtain such a model as well as a bound on the worst-case identification error.
\end{itemize} 
\end{problem}
The consistency is determined by using Carath$\hat{e}$odory-Fej$\acute{e}$r Interpolation~\cite{lim2003caratheodory}, defined below:
\begin{problem}
Given sequence of complex points, $g_i, i =0,1,2\cdots N-1 $, determine a function $G(z)$ such that 
\begin{equation}
    G(z)=g_{0}+g_{1}z^{1}+g_{2}z^{2}+\cdots +g_{N-1}z^{N-1}+\hat{g}_{N}z^{N}
\end{equation}
\end{problem}
The solution of this problem consists on following inequality. 
\begin{equation}
    I-T^{*}_{g}T_{g} \geq 0,
\end{equation}
where $T_{g}$ it the $n$x$n$ lower triangular Toeplitz matrix of sequence $\left \{ g_{0},g_{1},\cdots g_{n-1} \right \}$. The solution of Carath$\hat{e}$odory-Fej$\acute{e}$r Interpolation exits only if vector $g=g_{0}+g_{1}z^{1}+g_{2}z^{2}+\cdots +g_{N-1}z^{N-1}$ exists such that following Linear Matrix Inequality (LMI) holds:
\begin{equation*}
    M(g)=\begin{bmatrix}
KR^{-2} & (T^{N}_g)^{T}\\ 
 (T^{N}_g) & KR^{-2}
\end{bmatrix}\geq 0,
\end{equation*}
\begin{equation*}
    \left | y-T_u^N g\right |\leq \epsilon, 
\end{equation*}
where $R$ is the diagonal matrix with coefficients $\left \{ 1, \rho, \rho^{2}, \rho^{3} \cdots  \rho^{N-1} \right \}$. The $T_u^N$ and $T_g^N$ are the $n$x$n$ upper triangle matrix with coefficients $\left \{ u_{o}, u_{1}, u_{2}, \cdots  u_{N-1} \right \}$  and $\left \{ g_{o}, g_{1}, g_{2}, \cdots  g_{N-1} \right \}$, respectively. 

For non-parametric identification, the technique gives hard conservative results. To get less conservative results, the information on the parametric portion can also be included as priori information.  Assuming the non-parametric portion belongs to $\tau$, defined as:
\begin{equation}
    \tau\doteq P(z)=p^{T}G_{p}(z), p \in \Re ^{Np}
\end{equation}
where $p \in \Re ^{Np}$ is unknown vector of some  prior known component $G_p$ and $Np$ is the number of unknown parameters to be determined. This does not change the LMI problem defined above. It only changes the error bound defined as below:
\begin{equation*}
     \left | y-T_{u}^{N}pP+T_u^N g\right |\leq \epsilon, 
\end{equation*}
The $G(z)$ can be obtained by following set of equations~\cite{inanc2001robust}: 
\begin{gather*}
    G(z)= C_{G}(zI-A_{G})^{-1}B_{G}+D_{G}\\
A_{G}=A-[C_{-}^{T}C_{-}+(A^{T}-I)M_{R}]^{-1}C_{-}^{T}C_{-}(A-I)\\
B_{G}=-[(A^{T}-I)M_{R}+C_{-}^{T}C_{-}]^{-1}C_{-}\\
C_{G}=C_{+}[(A^{T}-I)M_{R}+C_{-}^{T}C_{-}]^{-1}C_{-}^{T}C_{-}(A-I)-C_{+}(A-I)\\
D_{G}=C_{+}[(A^{T}-I)M_{R}+C_{-}^{T}C_{-}]^{-1}C_{-}^{T}\\
A=\begin{bmatrix}
 0& I_{NxN} \\ 
 0& 0
\end{bmatrix}\\
C_=\overset{N}{\overbrace{\begin{bmatrix}
1 & 0  & \cdots & 0 
\end{bmatrix}}}\\
C_{+}=g^{T}R/K
\end{gather*}
In \emph{$l_{1}$} robust identification, the initial conditions are considered as zero, which is suitable for non-parametric estimation. However, to be able to include the effect of initial conditions in the system model the semi-blind robust identification technique is introduced in the Section-~\ref{sec:semi}.

\section{Semi-Blind System Identification}
\label{sec:semi}
The response of dynamic systems is highly affected by its state at  $t=0$. To include the initial condition effects, the identification technique described in Section~\ref{sec:l1} needs to be improved. The semi-blind robust identification technique takes into account information about the initial conditions of the system~\cite{ma2005semi}. The problem for semi-blind identification can be defined as follows: 
\begin{problem}
Given input sequence $u$, output sequence $y$, noise bound $\in \mathcal{N}$, maximum stability gain and characteristics of past input $u^{-}$, determine 
  $G(z)= G_{p}(z)+ G_{np}(z)$ which is compatible with priori and posteriori information, such that $\tau$ is non-empty set.
    \begin{equation}
    \tau (y)\doteq y_{i}=\sum_{i=0}^{N}g_{i}u_{N-i}+C_{g}A_{g}^{N-1}(\Gamma_{g}^{N}u^{-})_{i=0}
    \label{eq:semi-1}
    \end{equation}
    where $g_{0}=D_{g};g_{i}=C_{g}(A_{g})^{i-1}B_{g}$

\end{problem}
The solution to the above equation~(\ref{eq:semi-1}) involves solving a Bi-Affine Matrix, which is a non-convex, NP-hard problem. The above problem can be converted to the convex problem as mentioned in~\cite{ma2005semi}, which preserves the controllability and observability of the system. 
The convex problem can be defined as follows:
\begin{problem}

    Determine $G(z)= G_{p}(z)+ G_{np}(z)$, which is compatible with a priori and a posteriori information, such that $\tau$ is non-empty set:
    \begin{equation}
    \tau (y) =\left \{ G(z) \in S: y_{i} - (T_{g}^{N}u)_{i}+(\Gamma_{g}^{N}u^{-})_{i} \right \} 
\end{equation}
    where, $\left | (\Gamma_{g}^{N}u^{-})_{i} \right |\leqslant \gamma K_{u}; i=0,1, \cdots, N-1 $ and $T_{g}^{N}$ is the Toeplitz matrix and $\Gamma_{g}^{N}$ is the Hankel matrix.
\end{problem}

%where $T_{g}^{N}$ is the Toeplitz matrix and $\Gamma_{g}^{N}$ is the Hankel matrix. 
The first part of the $\tau$ set corresponds to the system response for input $u$ and the later part provides information for system response for past inputs $u^{-}$. This problem can be solved by following LMIs~\cite{ma2005semi}\cite{yilmaz2005robust}. 
 \begin{equation*}
    M(g)=\begin{bmatrix}
KR^{-2} & (T^{N}_g)^{T}\\ 
 (T^{N}_g) & KR^{-2}
\end{bmatrix}\geq 0,
\end{equation*}
\begin{equation*}
    \left | y-(T_u^N pP+T_u^N g)-\Gamma_{g}^{N}u^{-}\right |\in N, 
\end{equation*}
\begin{equation*}
    -\gamma K_{u} \leqslant \Gamma_{g}^{N}u^{-}\leqslant \gamma K_{u}
\end{equation*}
where $\gamma , K_{u}, p, P$ represent system gain, a bound $K_{u}$ on the norm of sequence $u^{-}$, affine parameters and the parametric portion of the system. 

\begin{figure}[H] \centering
    \centering
    \includegraphics[height=4cm]{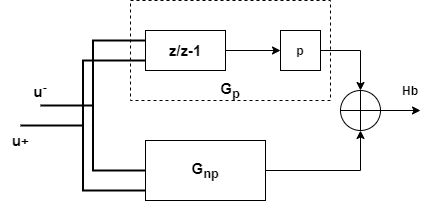}
    \caption{Framework for semi-blind identification.}
    \label{fig:param}
\end{figure}

The Fig.~\ref{fig:param} shows the framework for semi-blind identification. As the response to the EPO dosage as measurement of Hb is also effected by the past inputs $u^{-}$, the parametric part should be an integrator to accommodate all the effects of the inputs before $t=0$, i.e initial conditions.\\
As we have discussed the tools for parametric/non-parametric robust system identification to obtain a system model for the patients of chronic kidney disease for anemia management to control Hb level in response to EPO dosage. The Section-~\ref{sec:ARX} shows the implementation of \emph{$l_{1}$} robust identification and semi-blind robust identification on real patient data along with system validation using ARX one-step-ahead prediction method. 

\section{Simulation Results for Model Identification }
  \label{sec:ARX}
To obtained a model using identification techniques discussed in the above section, the EPO medication data and Hb measurement data of fifty patients have been obtained from the University of Louisville, Kidney Disease program. The EPO dosage is administrated three times a week and the Hb level is tested for only once a week. For system identification, the input/output data should be of the same length, hence three dosages of EPO per week are averaged which corresponds to a single Hb measurement. For model validation, the autoregressive model with exogenous output (ARX) is used~\cite{sekizawa2007modeling}, the difference equation for the ARX model is given as follows:
\begin{equation}
    a_{0}y(k)+a_{1}y(k-1)+a_{2}y(k-2)+\cdots+a_{n}y(k-n)=b_{1}u(k-n)+\cdots+b_{n}u(k-1)
\end{equation}
The models obtained by the \emph{$l_{1}$} robust identification and the semi-blind robust identification technique are compared by calculating minimum mean squared error (MMSE). The normalized MMSE mathematical definition in given in equation~(\ref{eq:MMSE})
\begin{equation}
    MMSE(k)=\frac{1}{N}\sum_{i=0}^{N}\frac{\left \| y(k)-\hat{y}(k) \right \|_{2}}{y(k)}
    \label{eq:MMSE}
\end{equation}

To obtain a robust model, the parameters $\rho = 1.01$, $\eta=0.31$ and $N=20$  are used as priori information for all patients. 

We present here the results for four patients, patient-1, patient-11, patient-21 and patient-32. The system model for patient number 1 obtained by \emph{$l_{1}$} robust identification and the ARX one step ahead prediction is shown below:
\begin{equation}
    G^{1}(z)=\frac{59.3 z^4 - 179.6 z^3 + 248.6 z^2 - 211 z + 92.98}{z^4 - 2.256 z^3 + 2.638 z^2 - 1.904 z + 0.6192}
\end{equation}

 \begin{figure}[H] \centering

    \subfigure[Model validation by ARX one-step ahead prediction for full and reduced order models along with MMSE analysis.]{\includegraphics[height=6cm, width=8cm]{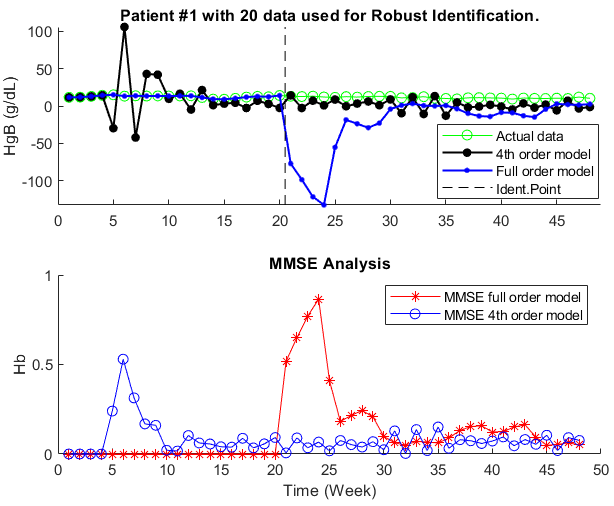}}
    \subfigure[Weekly EPO dosage and Hb data for Patient 1.]{\includegraphics[height=6cm, width=8cm]{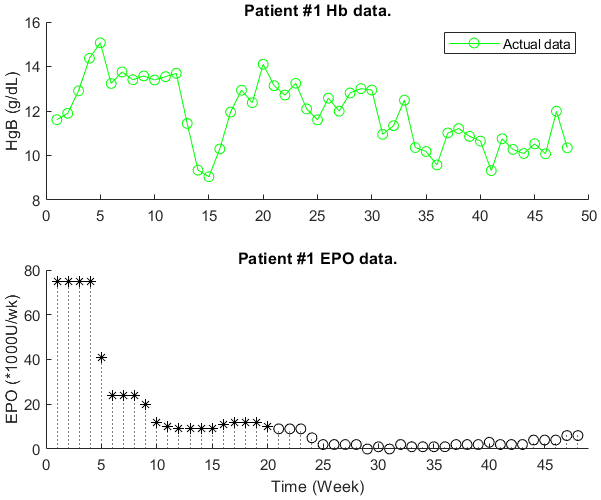}} 
    \caption{Responses and MMSE analysis of the patient-1 full and reduced order models obtained by the \emph{$l_{1}$} robust identification.}
    \label{fig:pat_1}
\end{figure}

The 20 samples of measurement of dose-response (EPO-Hb) data are used for the identification. After 20 samples the output is predicted by the ARX one-step-ahead prediction. As it is seen in Figure 3, the results of the \emph{$l_{1}$}-robust identification algorithm is poor. Even for the data points used in the identification (the first 20 data points) the output of the reduced order model misses the data points while the full order model matches the data points. However, for the data points (after the first 20 data points) which are not used in the identification both the full as well as reduced order models predictions do not match with the validation data. As a result, MMSE values as seen in Figure 3(a) are quite large. 

\begin{figure}[H] \centering
    \subfigure[Model validation by ARX one-step ahead prediction for full and reduced order models along with MMSE analysis.]{\includegraphics[height=5cm, width=8cm]{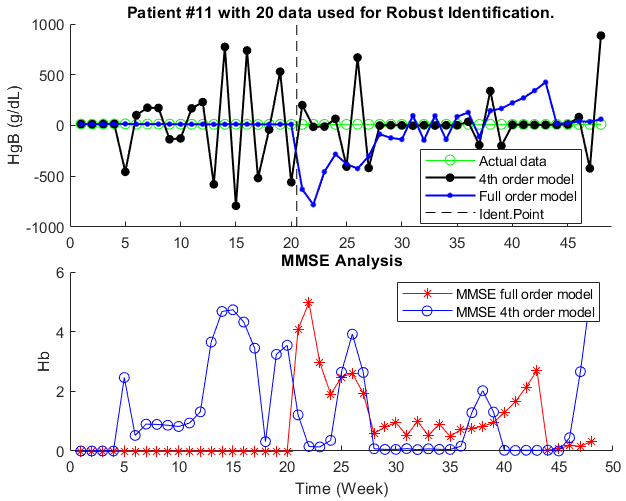}}
    \subfigure[Weekly EPO dosage and Hb data for Patient 11.]{\includegraphics[height=5cm, width=8cm]{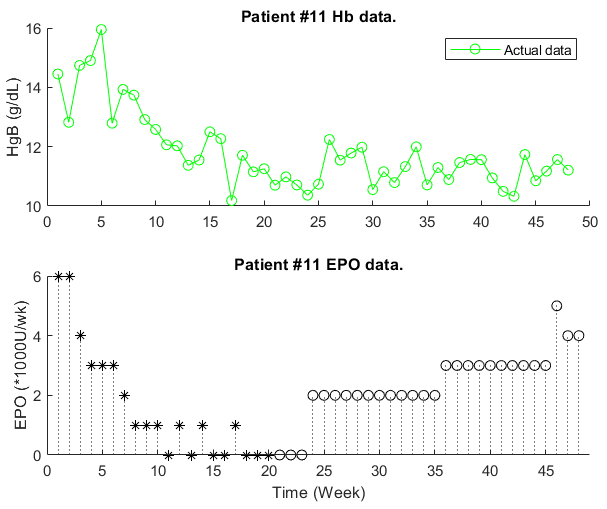}}   
    \caption{Responses and MMSE analysis of the patient-11 full and reduced order models obtained by the \emph{$l_{1}$} robust identification.}
    \label{fig:pat_11}
\end{figure}

\begin{figure}[H] \centering
    \subfigure[Model validation by ARX one-step ahead prediction for full and reduced order models along with MMSE analysis.]{\includegraphics[height=6cm, width=8cm]{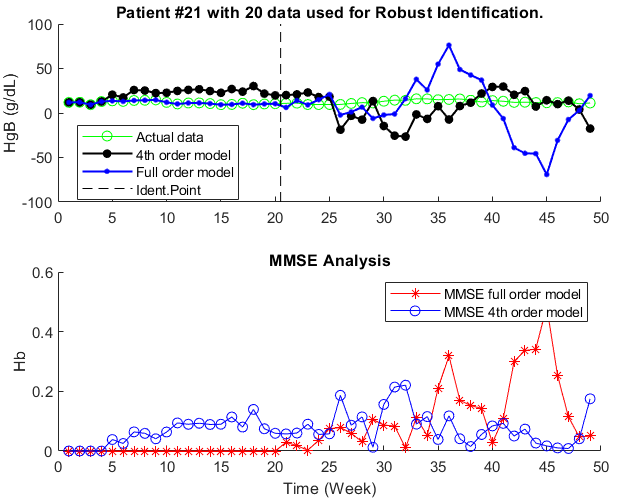}}
    \subfigure[Weekly EPO dosage and Hb data for Patient 21.]{\includegraphics[height=6cm,width=8cm]{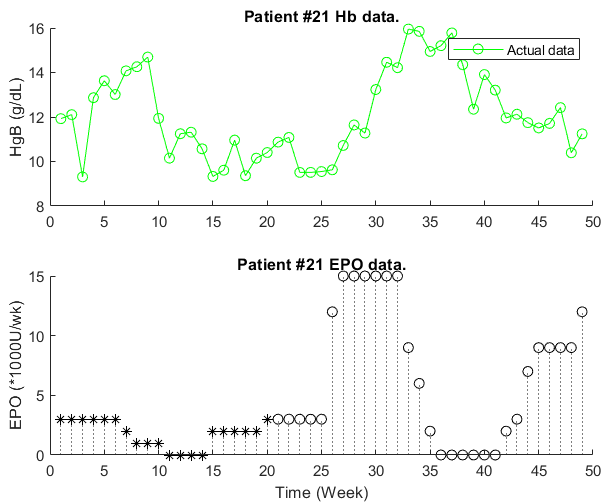}} 
    \caption{Responses and MMSE analysis of the patient-21 full and reduced order models obtained by the \emph{$l_{1}$} robust identification.}
    \label{fig:pat_21}
\end{figure}

Different than the patient-1, patient-11 dose-response data presents a more challenging case due to the several missing input (EPO) data for weeks 11,13,15,16,18 through 20. As before, the first 20 data points were used in the \emph{$l_{1}$}-robust identification algorithm to find the model. Then the full order model ($20^{th}$ order) is reduced to the $4^{th}$ order model. The full and reduced order models for patient-11 and model predictions are shown in Fig.~\ref{fig:pat_11}. 
\begin{equation}
    G^{11}(z)=\frac{   136.9 z^4 - 823.8 z^3 + 1879 z^2 - 1908 z + 739
}{ z^4 - 2.595 z^3 + 2.555 z^2 - 1.111 z + 0.1788}
\end{equation}
The model for patient number 21 and model validation is shown in Fig.~\ref{fig:pat_21}. The full and reduced order models for the patient-21 obtained by the semi-blind robust identification show low in the first 26 weeks of the data set as shown in MMSE analysis. The responses for full and reduced order models deviate from actual patient data as the EPO dosage is increased. 
\begin{equation}
    G^{21}(z)=\frac{-29.74 z^4 + 21.66 z^3 - 11.52 z^2 + 20.97 z - 30.32}{   z^4 - 0.7401 z^3 + 0.4182 z^2 - 0.7254 z + 0.9384}
\end{equation}

\begin{figure}[H] \centering
    \subfigure[Model validation by ARX one-step ahead prediction for full and reduced order models along with MMSE analysis.]{\includegraphics[height=6cm, width=8cm]{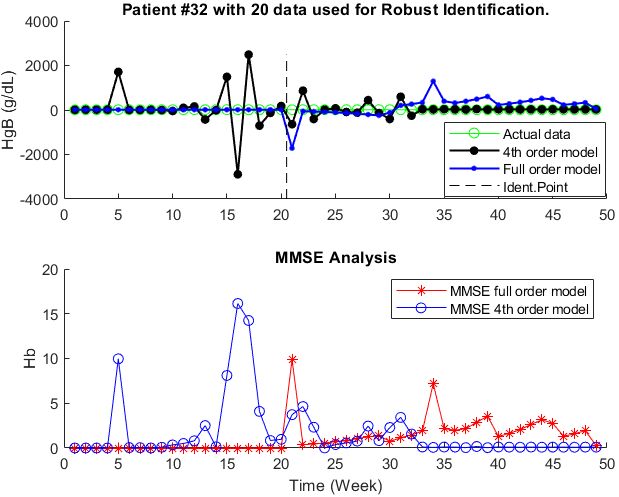}}
    \subfigure[Weekly EPO dosage and Hb data for Patient 32.]{\includegraphics[height=6cm, width=8cm]{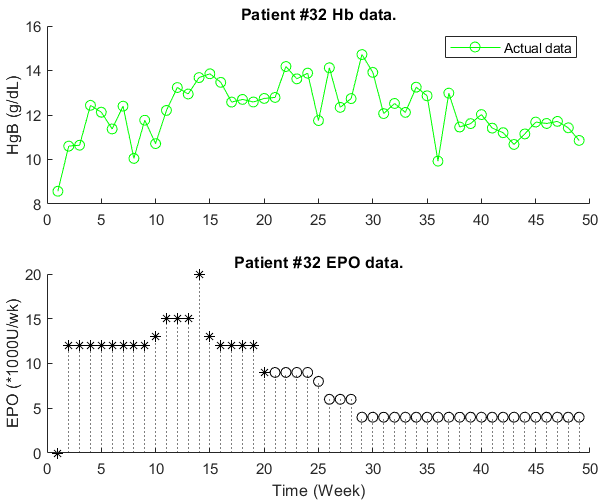}} 
    \caption{Responses and MMSE analysis of the patient-32 full and reduced order models obtained by the \emph{$l_{1}$} robust identification.}
    \label{fig:pat_32}
\end{figure}

The full order model is of $20^{th}$ order while the reduced-order model is $4^{th}$ order for all patients. The model for patient number 32 and model validation is shown in Fig.~\ref{fig:pat_32}:
\begin{equation}
    G^{32}(z)=\frac{ -584 z^4 + 2744 z^3 - 5031 z^2 + 4262 z - 1418}{z^4 - 3.027 z^3 + 3.592 z^2 - 1.968 z + 0.4198}
\end{equation}

\begin{figure}[H] \centering
    \subfigure[Model validation by ARX one-step ahead prediction for full and reduced order models.]{\includegraphics[height=6cm, width=8cm]{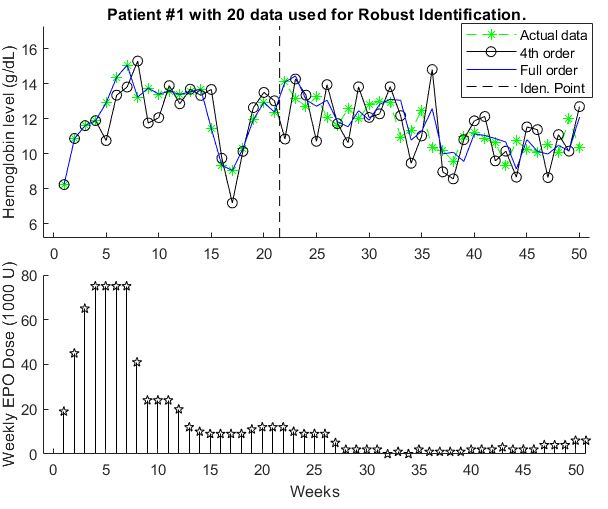}}
    \subfigure[MMSE for full order and reduced order models by ARX one-step ahead prediction]{\includegraphics[height=6cm, width=8cm]{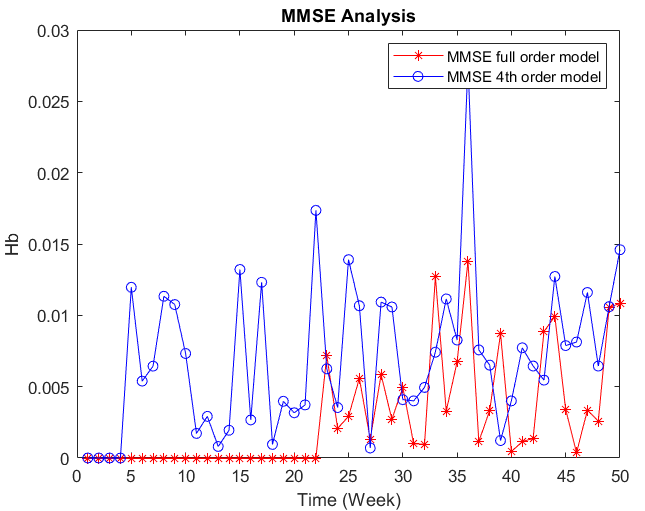}} 
    \caption{Responses and MMSE analysis of the patient-1 full and reduced order models obtained by the semi-blind robust identification.}
    \label{fig:pat_1_semi}
\end{figure}

These models assume the initial conditions equal to zero., which can highly affect the identified model as for the patients of Chronic Kidney disease (CKD), the past medication effects the state of the patient. Thus, these effects are included in the model using the semi-blind identification technique. The model of patient 1 obtained by the semi-blind identification technique and model validation results by ARX one-step-ahead prediction are shown in Fig.~\ref{fig:pat_1_semi}:  

\begin{equation}
    G^{1}(z)=\frac{1.307 z^4 - 0.9089 z^3 - 0.2374 z^2 + 0.9456 z + 0.03848}{z^4 - 1.649 z^3 + 0.4094 z^2 + 1.035 z - 0.7778}
\end{equation}

\begin{figure}[H] \centering
    \subfigure[Model validation by ARX one-step ahead prediction for full and reduced order models.]{\includegraphics[height=5cm, width=8cm]{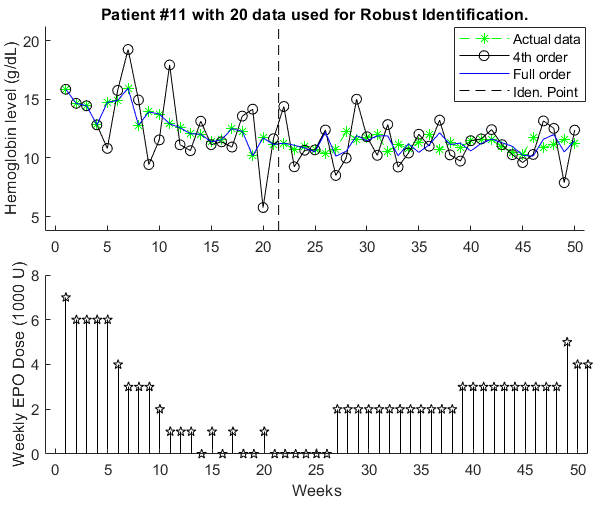}}
    \subfigure[MMSE for full order and reduced order models by ARX one-step ahead prediction.]{\includegraphics[height=5cm, width=8cm]{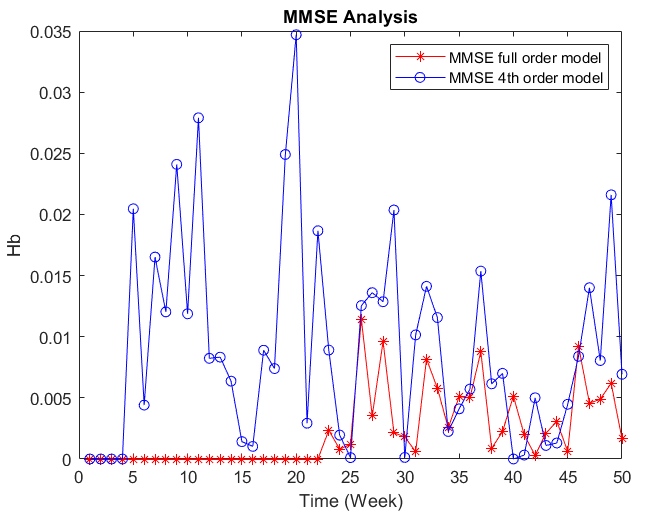}} 
    \caption{Responses and MMSE analysis of the patient-11 full and reduced order models obtained by the semi-blind robust identification.}
    \label{fig:pat_11_semi}
\end{figure}

The MMSE analysis for the patient-1 shows less error for the semi-blind robust identification as compared to the \emph{$l_{1}$} robust identification, as the semi-blind robust identification technique accounts the effect of the previous inputs. These results show that previous medication used by the patient significantly affects the model identification. The system model for patient number 11 obtained by semi-blind robust identification and model validation is shown in Fig.~\ref{fig:pat_11_semi}:
\begin{equation}
    G^{11}(z)=\frac{1.302 z^4 - 1.092 z^3 - 0.988 z^2 + 0.9426 z + 0.08707}{z^4 - 1.867 z^3 + 0.1278 z^2 + 1.542 z - 0.7983}
\end{equation}

\begin{figure}[H] \centering
    \subfigure[Model validation by ARX one-step ahead prediction for full and reduced order models.]{\includegraphics[height=6cm, width=8cm]{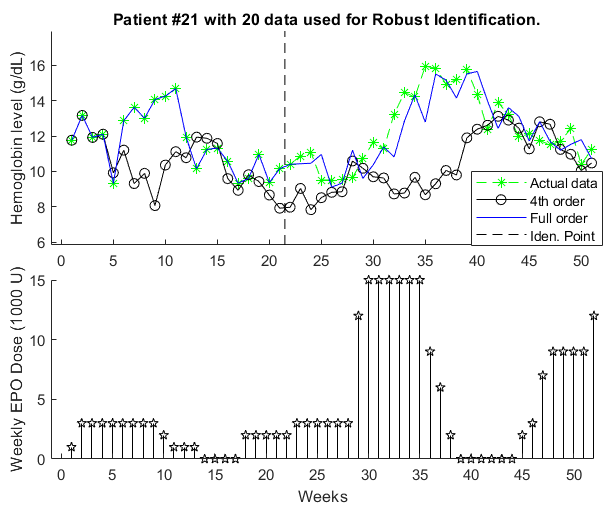}}
    \subfigure[MMSE for full order and reduced order model by ARX one-step ahead prediction.]{\includegraphics[height=6cm, width=8cm]{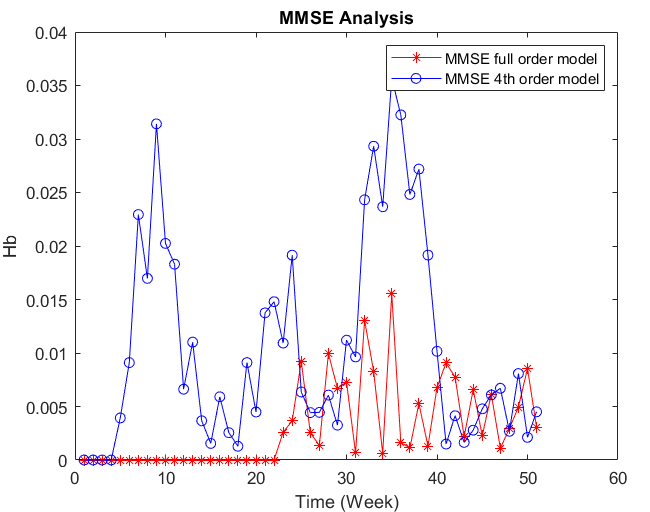}} 
    \caption{Responses and MMSE analysis of the patient-21 full and reduced order models obtained by the semi-blind robust identification.}
    \label{fig:pat_21_semi}
\end{figure}

The system model for patient number 21 obtained by semi-blind robust identification and model validation by the ARX one-step-ahead prediction is shown in Fig.~\ref{fig:pat_21_semi}:
\begin{equation}
    G^{21}(z)=\frac{1.417 z^4 + 1.594 z^3 + 1.217 z^2 + 1.157 z + 0.117}{z^4 + 0.1094 z^3 - 0.2046 z^2 - 0.07295 z - 0.7574}
\end{equation}

\begin{figure}[H] \centering
    \subfigure[Model validation by ARX one-step ahead prediction for full and reduced order models.]{\includegraphics[height=6cm, width=8cm]{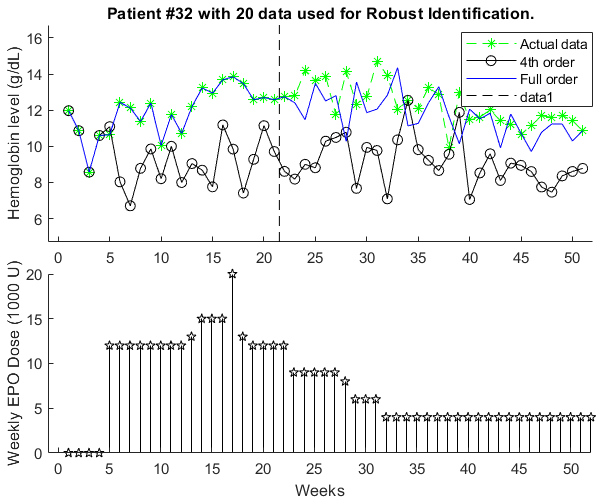}}
    \subfigure[MMSE for full order and reduced order model by ARX one-step ahead prediction.]{\includegraphics[height=6cm, width=8cm]{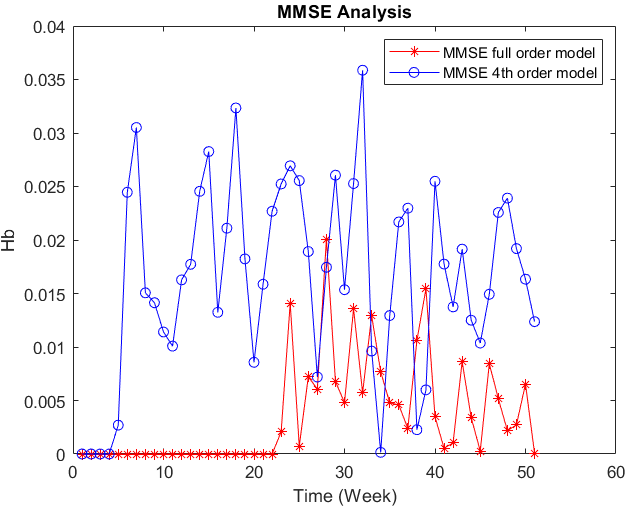}} 
    \caption{Responses and MMSE analysis of the patient-32 full and reduced order models obtained by the semi-blind robust identification.}
    \label{fig:pat_32_semi}
\end{figure}

The $4^{th}$ order model response for the patient-21 shows higher MMSE analysis for weeks 29 through 40. In this duration, the EPO dosage has increased suddenly which introduces a disturbance in the system and increases error.   The system model for patient number 32 obtained by semi-blind robust identification and model validation by the ARX one-steap ahead prediction is shown in Fig.~\ref{fig:pat_32_semi}:
\begin{equation}
    G^{32}(z)=\frac{2.744 z^4 + 4.082 z^3 + 1.548 z^2 + 0.0005763 z + 0.3473}{z^4 + 0.6091 z^3 - 0.8802 z^2 - 0.742 z + 0.07692}
\end{equation}

The reduced-order model for the patient-32 shows the high error, it is mainly due to the higher values of EPO dosage as the plot of EPO dosage shows continuous higher values. But the response of the full order model is very close to the actual patient data, which shows that the increase in a model can further reduce the error for the reduced-order model. By the MMSE analysis of full order model and reduced-order models obtained by semi-blind identification and \emph{$l_{1}$} robust identification technique, it is shown that semi-blind provides better results as it includes the effect of the past effect of medication.

\section{Conclusion}
  \label{sec:conclusion}
In this research work, the individualized system model for patients of chronic kidney disease (CKD) is obtained to regulate the external human recombinant erythropoietin (EPO) dosage to maintain the hemoglobin (Hb) levels between 11 and 12 $g/dL$. The \emph{$l_{1}$} robust identification and semi-blind identification technique have been implemented on the data set of fifty real patients obtained from the University of Louisville, Kidney Disease program. The \emph{$l_{1}$} robust identification is non-parametric identification which considers the zero initial conditions. However, the semi-blind identification technique determines the effects of inputs before  $t=0$ as well as the system model. Both of these techniques do not assume the probabilistic distribution of noise or predefined model structure. The obtained patient models are validated by using the ARX one-step prediction method. The minimum mean square error (MMSE) is calculated for each patient between actual data and predicted data. The error in the model obtained in semi-blind identification techniques is less than \emph{$l_{1}$} robust identification technique. This research work provides a basis to obtain individualized models for new patients. The next phase is to design the controller for the patient based on the identified models and automate this procedure for new patients.

\section{Acknowledgments}
  \label{sec:Acknowledgments}
 This work was supported by NSF under grant 1722825.

%\section*{References}
\bibliographystyle{elsarticle-num}
\bibliography{reference}

%\begin{thebibliography}{00}

%% \bibitem[Author(year)]{label}
%% Text of bibliographic item

%\bibitem[ ()]{}

%\end{thebibliography}

\end{document}